\documentstyle[sprocl]{article}
\begin{document}
\title{SCATTERING OF ULTRAHIGH ENERGY (UHE) EXTRAGALACTIC NEUTRINOS ONTO LIGHT
RELIC NEUTRINOS IN GALACTIC HDM HALO OVERCOMING THE GZK CUT OFF}

\author{D. Fargion, B. Mele}

\address{Physics Dept. Rome University 1 and INFN,\\
Piazzale A.Moro 2, 00185 Rome, Italy}

 \maketitle

\begin{abstract}
The rarest cosmic rays above the GZK cut-off $(E_{CR} \tilde{>}
10^{19} \div 10^{20} eV)$ are probably born at cosmic distances
($\tilde{>}$ tens Mpc) by AGNs (QSRs, BLac, Blazars...). Their
puzzling survival over $2.75 K^o$ BBR radio-waves opacities (the
``GZK cut off'') might find a natural explanation if the traveling
primordial cosmic rays were UHE neutrinos (born by UHE
photopion decay) which are transparent to $\gamma$ or $\nu$ BBR. These UHE $%
\nu$ might scatter onto those (light and cosmological) relic neutrinos
clustered around our galactic halo.\newline
The branched chain reactions from a primordial nucleon (via photoproduction
of pions and decay to UHE neutrinos) toward the consequent beam dump
scattering on galactic relic neutrinos is at least three order of magnitude
more efficient than any known neutrino interactions with Earth atmosphere or
direct nucleon propagation. Therefore the rarest cosmic rays (as the 320 EeV
event) might be originated at far $(\tilde{>} 100 Mpc)$ distances (as
Seyfert galaxy MCG 8-11-11). The needed UHE radiation power is in rough
agreement with the NCG 8-11-11 observed in MeV gamma energy total output
power. The final chain products observed on Earth by the Fly's Eye detector
might be mainly neutron and antineutrons as well as, at later stages,
protons and antiprotons. These hadronic products are most probably
secondaries of $W^+ W^-$ or $ZZ$ pair productions and might be consistent
with the last AGASA discoveries of 6 doublet and one triplet event.
\end{abstract}

\bigskip
\bigskip
\section{Introduction}
\bigskip

Most energetic cosmic rays UHE $(E_{CR} > 10^{19} eV)$ are bounded to short $%
(\tilde{<} 10 Mpc)$ distances by the 2.73 K BBR opacity [1], [2] (the GZK
cut off) and by diffused radio intergalactic noise [3], [4]. The main
electromagnetic ``viscosities'' stopping the UHE cosmic ray (nuclei,
nucleons, photons, electrons) propagation above $\sim 10 Mpc$ are:

\noindent (1) The Inverse Compton scattering of any charged lepton (mainly
electrons) on the BBR $(e^\pm_{CR} \gamma_{BBR} \rightarrow e^\pm \gamma)$
[5], [6] [7],

\noindent (2) the nucleons photopair production at higher energies $(p_{CR}
+ \gamma_{BBR} \rightarrow p e^+ e^-)$,

\noindent (3) the UHE photon BBR or radio photon electron-pair productions $%
(\gamma_{CR} + \gamma_{BBR} \rightarrow e^+ e^-)$,

\noindent (4) nuclei fragmentation by photopion interactions.

\noindent (5) the dominant nucleon photoproduction of pions $(p_{CR} +
\gamma_{BBR} ~\rightarrow ~ p + N \pi ; n + \gamma_{BBR} \rightarrow n + N
\pi)$.

The above GZK constrains apply to all known particle (proton,
neutron, photons, nuclei) excluding neutrinos. Nevertheless, the
UHE cosmic rays, either charged or neutral, flight straight
keeping memory of the primordial source direction, because of the
extreme magnetic rigidity [8]. However all last well localized
most energetic cosmic rays (as the Fly's Eye 320 EeV event on
October 1991) do not exhibit any cosmic nearby $(< 60 Mpc)$ source
candidate in the same arrival direction error box. From here the
UHE cosmic ray paradox arises. The few known solutions are
difficult to accept:

\noindent (a) An {\it exceptional} $(B \tilde{>} 10^{-7} Gauss)$ {\it %
coherent} magnetic field on huge extragalactic distances able to
bend (by a large angle) the UHECR trajectory coming not from
distant but from nearer off-axis sources like M 82 or Virgo A [9].

This solution was not found plausible [10].

\noindent (b) Exotic topological defect annihilations [9] in
diffused galactic halo is an ad hoc, and a posteriori solution.
Moreover it is in contradiction with recent evidences by AGASA
detector of data inhomogeneities, i.e. of doublets or triplet
UHECR events arriving from the same directions [12].

\noindent (c) A galactic halo population of UHECR sources (as the
fast running pulsars associated with SGRs[11]). These small size
(neutron star - black-hole) jets sources must be extremely
efficient in cosmic ray acceleration at energies well above the
expected common maximum energy $E < BR \simeq \tilde{<} 10^{17} eV
({\frac{B}{3.10^{-6} G}}) ({\frac{R}{50pc}})$ required by a
supernova accelerating blast wave. Moreover their extended halo
distribution must reflect in a dipole and/or in a quadrupole UHECR
anisotropy, signature not yet identified. Therefore it seems at
least premature to call for a solution in (local galactic - local
group) location of microquasar jets.

\noindent (d) A direct nucleus or a nucleon at high energies will
be at 100Mpc distances severely suppressed by GZK $(10^{-5})$ cut
off and, more dramatically, they will induce a strong signal of
secondaries at energies just below the GZK bound, a totally
unobserved signal.

Our present [13,14] solution is based on the key role of light
$(m_\nu \tilde{>} eV)$ cosmic neutrinos clustered in extended
galactic halo: these relic neutrinos act as a target calorimeter
able to absorb the UHE $\nu$'s from cosmic distances and to
produce hadronic showers in our galaxy. The primary UHECRs are the
usual AGNs or Blazars able to produce huge powers and energies.
Their photopion production and decays, near the source, into
muonic and electronic neutrinos, generate the main $\nu$s
messenger toward cosmic distances up to our galactic halo. Their
final interactions with clustered relic $\nu_r$ (and
$\bar{\nu}_r$) of all flavours (but
preferentially with the heaviest and best clustered one $(\nu_\tau , \bar{\nu%
}_\tau)$) may offer different channel reactions:

\noindent (A) $\nu \bar{\nu}_\tau$ scattering via a Z exchanged in
the s-channel leading to nucleons and photons.

\noindent (B) $\nu \bar{\nu}_r$ scattering via t-channel of
virtual W exchange between different flavours. This is able to
produce copious UHE
photons (mainly by $\nu_\mu \bar{\nu}_\tau \rightarrow \mu^- \tau^+$ and $%
\tau$ pion decay),

\noindent (C) $\nu \bar{\nu}_r$ production of $W^- W^+$ or ZZ pairs. The
latter channels are the best ones in our opinion to produce final nucleons $%
(p, \bar{p} , n ,\bar{n})$ which fit observational data.

\vskip 12pt \noindent {\bf The UHE $\nu$ cross section interacting with
relic $\nu_\tau, \bar{\nu}_\tau$.} \vskip 6pt

The general framework to solve the GZK puzzle we proposed is a tale story
beginning from a far AGN source whose UHE protons $(E_p \tilde{>} 10^{23} eV)
$ are themselves a source of pions and secondaries muons and UHE neutrinos.
The latter may actually escape the GZK cut off, traveling unbounded all the
needed cosmic distances $(\sim 100 Mpc)$.

Once near our galactic halo, the denser gravitationally-clustered
light relic neutrinos, forming a hot dark halo, might be able to
convert the UHE $\nu$'s energies by scattering and subsequent
decays into an observable nucleon (or antinucleon), the final
observed UHECR remnant. The $\nu - \nu$ interaction cross-sections
are the key filter which make possible and efficient the whole
process. In fig. 1 from [14] we show the three main processes
cross-section as a function of the center of mass energies. The (s
- channel)
$\nu_\mu \bar{\nu}_{\mu R} \rightarrow Z$ exhibits a resonance at $%
E_\nu = 10^{21} eV ({\frac{m\nu}{4eV}})^{-1}$; (t - channel)
$\nu_\mu \nu_{\tau_R} \rightarrow \mu^- \tau^+$ via virtual W.
These reactions are the most probable ones but UHE photon seem to
be excluded by geomagnetic high height cut off. The $\nu
\bar{\nu}_\tau \rightarrow W^+ W^-$ cross section
is also shown. There is an additional Z pair production channel $(\nu \bar{%
\nu}_\tau \rightarrow ZZ)$ [17] almost coincident in its general
behaviour with the $W^+ W^-$ production. It is not shown on the
figure. Its global contribute (to be discussed elsewhere) is to
double the $(\nu \bar{\nu}_\tau \rightarrow W^- W^+)$ chain
products making easier their detection [17]. The table 5 from
[14], (regarding the $W^+ W^-$ channel) is also shown. It
describes the complex reaction branching (left column) of the
reaction considered, the corresponding probability (left-center
column ), the consequent multiplicity (center - right column of
the by-products), the final energy of their secondaries. For
example, the first reaction shows the nucleon $+ \gamma_{BBR}$
photopion production at the primordial proton/neutron energy (see
the final box $E_p^{WW} \cong 7.10^{23} eV$), the corresponding
multiplicity, $8 \pi$, which is reduced to 5 for the charged
pions, and the corresponding relativistic, secondary pion energy
$E_\pi \sim {\frac{Ep}{9}}$.

The following pion and muon decay in UHE neutrinos are
straightforward. The $\nu_\mu \bar{\nu}_{\mu \tau} \rightarrow W^+
W^-$ reaction and probability must be considered at this energy
range assuming a neutrino density clustering [15,16] in galactic
halo: $P \sim \sigma n_{\nu _\tau} l_g$.

The clustered neutrino density contrast [15,16] is comparable to
the barionic one ${\frac{n\nu_\tau}{n \nu_{BBR}}} \sim
{\frac{\rho_G}{\rho_W}} \sim 10^{5\div 7}$. In conclusion [14],
the total probability of the processes and the corresponding
needed primordial proton energy $E_p^{WW}$ (calibrated by the
final observed nucleon cosmic ray at 320 EeV = $4.3 \cdot 10^{-4}
E_p^{WW}$) are summarized in the square boxes. The probability
(taking into account global multiplicity and probability) to occur
is at least $P^{WW} \tilde{>} 10^{-3} ({\frac{m\nu}{10eV}})^{-1}$
corresponding, for the candidate source MCG 8-11-11, to a needed
average power $E^{WW} \sim 2.5 \cdot 10^{48} ergs^{-1}$. The
additional $\nu \bar{\nu} \rightarrow ZZ$ channel [17] will reduce
to half the power above [14], making the value comparable with the
MCG8-11-11 observed low-gamma MeV luminosity $(L_\gamma \sim 7
\cdot 10^{46} erg s^{-1})$. We predicted parassite signals photons
at $10^{16} eV$ energies [14] as well as a peculiar imprint on
larger sample data, due to the central overlapping of
neutron-antineutron prompt arrival toward the source line of
sight. An additional twin mirror (deviated) signal due to protons
and antiprotons random walk will arrive late nearly at opposite
(few degrees) sides. These characteristic signatures might be
already recorded by AGASA in the last few doublet and triplet
UHECR events.

The UHE neutrinos above the GZK cut-off are observable from almost
all the Universe while the corresponding UHE nucleons (or gamma)
above the GZK energies are born in a smaller constrained ``GZK''
volume.Therefore the expected flux ratio for UHE $\nu$'s over
nuclei or photons at GZK energies is roughly (in euclidean
approximation) independent of the source spectra:

\[
{\frac{\phi_\nu }{\phi_{GZK}}} \sim [{\frac{z_\nu }{z_{GZK}}}]^{3/2} \simeq
3 \cdot 10^4 \eta ({\frac{z_\nu}{2}})^{3/2} ({\frac{z_{GzK}}{2 \cdot 10^{-3}}%
})^{-3/2}
\]

where $z_\nu$ is a characteristic UHE $\nu^{\prime}$s source redshift $%
\simeq 2$ and $z_{GZK} \sim 2\cdot 10^{-3}$. Assuming an efficiency ratio $%
\eta$ for the conversion from UHE proton to UHE $\nu$'s of a few percent,
the ratio $\phi_\nu /\phi_{GZK} \tilde{>} 10^3$ is naturally consistent with
the inverse probability $(P^{WW} \tilde{>} 10^{-3})^{-1}$ found above.

Therefore a $10^{+3}$ fold larger flux of UHE $\nu$'s (than the
corresponding nucleon flux C.R.) above GZK (mainly of tau nature [18])
should be observed easily in a $Km^3$ neutrino detector in a very near
future.

\end{document}